\documentclass{an}
\usepackage{graphicx}
\usepackage{times}
\Volume{000}             % volume number
\Year{0000}              % format: {YYYY}
\Month{00}               % format: {M} or {MM} (number)
\Pagespan{000}{999}      % format: {start page}{last page}

\begin{document}
\lhead[\thepage]{M. Wittkowski et al.: Measuring starspots on magnetically 
active stars with the
  VLTI} \rhead[Astron. Nachr./AN~{\bf XXX} (200X)
X]{\thepage} \headnote{Astron. Nachr./AN {\bf XXX} (200X) X, XXX--XXX}

\title{Measuring starspots on magnetically active stars with the VLTI}

\author{M. Wittkowski\inst{1} \and M. Sch\"oller\inst{1} \and 
S. Hubrig\inst{1} \and B. Posselt\inst{2} \and O. von der L\"uhe\inst{3}} 
\institute{European Southern Observatory, Casilla
  19001, Santiago 19, Chile \and 
Astrophysikalisches Institut und Universit\"ats-Sternwarte,
%Friedrich Schiller Universit\"at Jena,
Schillerg\"a{\ss}chen 2-3,
07745 Jena, Germany
\and Kiepenheuer-Institut f\"ur
  Sonnenphysik, Sch\"oneckstr. 6, 79104 Freiburg, Germany}
%Comment: Adresse fuer Uni Jena fehlt noch

\date{Received {\it date will be inserted by the editor}; accepted
  {\it date will be inserted by the editor}}

\abstract{ We present feasibility studies to directly image stellar  
  surface features, which are caused by magnetic activity, with the
  Very Large Telescope Interferometer (VLTI).  We concentrate 
  on late type magnetically active stars, for which the distribution 
  of starspots on the surface has been inferred from photometric and 
  spectroscopic imaging analysis. The study of the surface spot evolution 
  during consecutive rotation cycles will allow first direct measurements 
  (apart from the Sun) of differential rotation which is the central 
  ingredient of magnetic dynamo processes. The VLTI will provide baselines 
  of up to 200 m, and two scientific instruments for interferometric studies 
  at near- and mid-infrared wavelengths. Imaging capabilities will be made 
  possible by closure-phase techniques. We conclude that a realistically 
  modeled cool surface spot can be detected on stars with angular
  diameters exceeding $\sim$\,2\,mas using the VLTI with the first
  generation instrument AMBER. The spot parameters can then be derived with
  reasonable accuracy. We discuss that the lack of knowledge of magnetically
  active stars of the required angular size, especially in the southern 
  hemisphere, is a current limitation for VLTI observations of these 
  surface features.  
 }

\correspondence{Markus Wittkowski, mwittkow@eso.org}

\maketitle

\section{Introduction}
Optical interferometers are about to become powerful instruments to
image vertical and horizontal temperature profiles and inhomogeneities 
of stellar surfaces. 
Direct observations of these structures
will be the key for constraining underlying hydrodynamic and 
magneto-hydrodynamic mechanisms and for our further understanding 
of related phenomena like predictions of stellar activity and 
asymmetric or stochastic mass-loss events. 
The study of the evolution of surface spots which are caused by magnetic 
activity, during consecutive rotation cycles will allow first direct 
measurements (apart from the Sun) of differential rotation which is 
the central ingredient of magnetic dynamo processes.

Many starspots have been discovered by photometric monitoring 
(e.g. Strassmeier et al. 1999). The distribution of
starspots on the surfaces of stars has so far been inferred from
photometric and spectroscopic imaging analysis for about 50 stars
(''Summary of Doppler images of stars'', 
www.aip.de/groups/activity/DI/summary). In addition, there is evidence
for surface features on some slowly rotating single K giant
stars based on Ca\,II variability (e.g. Choi et al. 1995). The observation of
these stars may not be very feasible for Doppler imaging techniques
because of their slow rotation. However, they might be good candidates for 
interferometric imaging, here because of their slow rotation, and their
sufficiently large angular diameters (Hatzes et al. 1997). Magnetic activity 
is usually assumed to be the cause for these surface spots.

Optical interferometry has already proven its ability to derive
stellar surface structure parameters beyond diameters.  The
limb-darkening effect on stellar disks was confirmed by
interferometry for several stars by Hanbury Brown et al. (1974),
Haniff et al. (1995), Quirrenbach et al. (1996), Burns et al. (1997),
Hajian et al. (1998), Young et al. (2000), and Wittkowski et al.
(2001). The latter measurement did not only succeed in discriminating
uniform disks and limb-darkened disks, but also in constraining model
atmosphere parameters. For the apparently largest supergiants
$\alpha$~Orionis, $\alpha$~Scorpii and $\alpha$~Herculi, bright spots 
have been detected and mapped by direct imaging techniques including optical
interferometry and HST imaging (Buscher et al. 1990, Wilson et al.
1992, Gilliland \& Dupree 1996, Burns et al. 1997, Tuthill et al.
1997, Young et al. 2000). Large-scale photospheric convection 
(Schwarzschild 1975, Freytag et al. 1997) is a preferred interpretation 
for the surface features on the these supergiants, and for interferometric 
observations of asymmetric features on or near stellar disks 
(see, e.g. Perrin et al. 1999, Weigelt et al. 2000) or circumstellar 
envelopes (see, e.g. Weigelt et al. 1998, Wittkowski et al. 1998).

The ESO VLTI (Very Large Telescope Interferometer), a European general 
user interferometer located on Cerro Paranal in northern Chile, 
is about to become an excellent facility to study stellar surfaces by 
means of optical interferometry. It will provide a large choice of 
baselines up to 200\,m, the use of the 8\,m Unit Telescopes 
(maximum baseline 130\,m), of several 1.8\,m Auxiliary Telescopes as well 
as of a near-infrared instrument (AMBER, see Petrov et al. 2000), 
and a mid-infrared instrument (MIDI, see Leinert et al. 2000). 
The AMBER instrument will provide imaging capabilities by
closure-phase techniques.
For a description of the VLTI and the recent achievement of first fringes 
see Glindemann et al. (2000), Glindemann et al. (2001a),
Glindemann et al. (2001b), and references therein. 
Since then, commissioning activities have been carried out.
VLTI commissioning data of scientific relevance are being publicly
released 
({\small www.eso.org/projects/vlti/instru/vinci/vinci\_data\_sets.html}).
Measurements of visibility amplitudes beyond the first minimum
of the visibility function, which is mandatory for an analysis of
stellar surface features, became already feasible with the commissioning
instrument VINCI
(see ESO Press Release 23/01; Wittkowski et al., in preparation). 
The potential to study stellar surfaces with the VLTI has previously 
been investigated
by von der L\"uhe et al. (1996), Hatzes (1997) and von der L\"uhe
(1997). Other existing or planned ground-based optical and near-infrared
interferometric imaging facilities (with location and maximum baseline)
include the CHARA (Mt. Wilson, USA, 350\,m), COAST (Cambridge, UK, 100\,m),
IOTA (Mt. Hopkins, USA, 40\,m), NPOI (Flagstaff, USA, 437\,m), KECK 
(M. Kea, USA, 140\,m), LBT (Mt. Graham, USA, 22\,m), 
OHANA (M. Kea, USA, 800\,m). In addition, free-flying space-based 
interferometers are being designed. For instance, the ``stellar imager'' (SI),
a NASA design for a space-based UV/optical interferometer with a 
characteristic spatial resolution of 0.1 mas, is especially intended to 
image dynamo activity of nearby stars (hires.gsfc.nasa.gov/~si).

Here, we investigate to which extent and accuracy surface features with
realistically modeled parameters can be analyzed in the near future
by VLTI measurements using first generation instruments. 
Observing restrictions and accuracy limits are taken into account. 
We concentrate on magnetically active stars and studies which 
aim at understanding stellar magnetic dynamo processes.
\section{Potential sources}
The VLTI with its first generation instruments will provide a maximum
baseline $B$ of 200\,m and a shortest wavelength $\lambda$ of
1.0\,$\mu$m using the AMBER instrument at the lower end of its 
wavelength range.  
With this instrument, one resolution element $\lambda/B$ will correspond 
to an angular size of 1.0\,mas. 
Requiring at least two resolution elements across the
stellar disk, the minimum diameter is about 2.0\,mas. 
The use of the longest (200\,m) VLTI baseline implies that the 
1.8\,m Auxiliary Telescopes are employed, and the anticipated limiting 
J magnitude is about 6\,mag, assuming that fringes are tracked 
on a shorter baseline (baseline bootstrapping).

However, despite these limitations for the use of the VLTI,
other ground-based optical interferometers, as for instance the completed
NPOI with a characteristic maximum resolution of $\sim$\,0.3\,mas, or even
planned space-based UV interferometers (e.g. SI), will be able to 
image smaller bright stars.
Consequently, the search for potential sources, i.e. for
stars which are known to exhibit spots, was not restricted due to diameters.
Stellar diameters were estimated for stars which are known or expected
to have spots on their surface. 
The resulting full table is described and shown
in Appendix~A. This list serves as source of information
on potential targets for interferometric imaging.
Depending on the location and characteristics (e.g. baseline, 
observing wavelengths, limiting magnitude) of an interferometer, 
certain restrictions apply to the choice of feasible targets from
the lists in Appendix~A. Restrictions due to the array geometry like 
shadowing effects or limited availability of delay line lengths might 
arise in addition. 
Further limitations on the amount of observational data that can 
be combined for the purpose of imaging techniques may apply due to 
the rotational velocity of the star. 

Most stars on the lists in Appendix~A belong to one of the following
classes of objects:
RS\,CVn type, BY\,Dra type, W\,UMa type, FK\,Com type, T\,Tau type
and single giants. Single giants are in principle good candidates 
for interferometric imaging, because they are relatively large and 
usually they rotate slowly.
In addition, observations are not affected by companions in the
field of view\footnote{The VLTI field of view is of the order of 2 arcsec 
without spatial filtering or, with spatial filtering, of the size of an 
Airy disk of a single telescope, i.e. $\sim$\,115 mas with the ATs at 
a wavelength of 1\,$\mu$m.}.
However, most single giants are not expected to be magnetically active and
information on parameters of spots caused by magnetic activity is rare.
Doppler images have so far been obtained for only one magnetically active 
effectively single giant, namely HD\,51066 or CM\,Cam
(Strassmeier et al. 1998). HD\,27536 is another example of this rare class
and has been discovered by photometric monitoring (Strassmeier et al. 1999b). 
Other single giants are suspected to be magnetically active and to 
harbor spots on their surfaces, based
on Ca\,II variability (Choi et al. 1995). 

\begin{table}
\caption{Stars with known surface spots from the Tables in the
Appendix which are feasible for VLTI observations with AMBER, 
i.e which have an estimated
diameter of larger than 2\,mas, a declination of less than 30\,degrees, and a 
J magnitude of brighter than 6. Shown are the star's name, variability type, 
declination, rotation period, and diameter. The diameter is the
mean of measured diameters, if available, or the mean of estimated diameters.
The last column indicates whether or not Doppler images are available.}
\label{tab:largediam}
\begin{tabular}{llrrrrr}
\hline
Name   & Type     & Dec.& P    & $\theta$ & Doppler \\
       &          &(deg)& (d)  & (mas)    & images \\
\hline
$\zeta$\,And    & RS\,CVn  & +24 & 17.6 & 2.8     & no \\
$\sigma$\,Gem   & RS\,CVn  & +28 & 19.5 & 2.4     & yes\\
$\Theta^1$\,Tau & giant    & +16 &      & 2.6     & no \\
$\gamma$\,Com   & giant    & +28 &      & 2.5     &  no \\
\hline
\end{tabular}
\end{table}
%Comment: Add errors for diameters!
%
Table \ref{tab:largediam} shows four stars from the lists in Appendix~A
which are feasible for VLTI observations. They include
two RS\,CVn type variables and two single giants. 
The rotational periods of the
RS\,CVn type stars $\zeta$\,And and $\sigma$\,Gem of 17.6 days and 
19.5 days, respectively, allow observations during 6 hours with 
a smearing effect of only about 5 degrees. However, the existence of 
the companion stars may lead to effects which complicate the interpretation
of the visibility data. Doppler images exist for the RS\,CVn type giant
$\sigma$\,Gem, providing us with characteristics on the spots.

The objects listed in the Appendix~A have mostly been observed in the
northern hemisphere.
It would be desirable to
initiate similar photometric and spectroscopic variability studies in 
the southern hemisphere which could serve as a preparation and 
source of complementary information for VLTI measurements of stellar surfaces.
\section{Interferometric imaging}
The observables of an interferometer are the amplitude and phase of the
complex visibility function, which is the Fourier transform of the
object intensity distribution. In presence of atmospheric turbulence, 
usually only the squared visibility amplitudes and the triple products
are accessible by optical 
and near-infrared interferometers. The triple product is the product
of three complex visibilities corresponding to baselines that form a 
triangle. The phase of the triple product, the closure phase, is free 
of atmospheric phase noise (Jennison 1958).
An image can usually only be reconstructed if at least 
three array elements are used simultaneously\footnote{An option with
two or more array elements is phase referenced imaging which measures 
the fringe phase relative to that of a reference object within the 
isoplanatic patch. For Michelson style interferometers, this method
will usually need a dual feed system and is, hence, very difficult 
to realize.}. 
The general methods to reconstruct an image from sparse aperture 
interferometric visibility data have been developed by radio astronomers. 
In fact, the first images (Baldwin et al. 1996, Benson et al. 1997, 
Hummel et al. 1998) that were reconstructed from optical interferometric 
data using the COAST and NPOI interferometers employed imaging software 
which is usually used by radio interferometrists. 
As an example for interferometric imaging of starspots, Hummel (1999) showed
by simulated NPOI data that a star with one spot (stellar diameter 12\,mas, 
spot diameter 3\,mas, $T_\mathrm{star}=5000$\,K, $T_\mathrm{spot}=3500$\,K) 
can be imaged by NPOI.
A recent review describing the image fidelity using optical interferometers 
can be found in Baldwin \& Haniff (2002). 
 
In order to reconstruct an image of reasonable fidelity, it is intuitively 
understandable that the number of visibility data has to be at least as 
large as the number of unknown variables, i.e. the pixel values of the image
(see, e.g. Baldwin \& Haniff 2002).
Furthermore, despite of the ability of imaging algorithms to effectively
interpolate the sparse aperture data, the aperture has to be filled
as evenly as possible. In case of highly unevenly filled apertures,
the point spread function would show strong artifacts (sidelobes) 
which could not be deconvolved. This implies that most probes of the
visibility function have to be taken at points beyond its first minimum.
%Comment klarer ausdruecken !
Here, the visibility amplitude is very low, corresponding to vanishing
fringe contrasts. This makes it difficult to detect and track the
fringes. As a solution, these low fringe contrasts can be recorded 
while fringes are tracked on shorter baselines with higher contrasts using 
bootstrapping techniques (see, e.g. Roddier 1988, Armstrong et al. 1998,
Hajian et al. 1998, Wittkowski et al. 2001).  

The VLTI with the AMBER instrument allows the simultaneous combination 
of three beams. The maximum feasible size of an image of a stellar disk 
is limited by the number of visibility data, which can be combined,
to less than about 10\,$\times$\,10\,pixels.
For the stars in Tab.~\ref{tab:largediam}, 
which have diameters of only $\sim$\,2.5\,mas, the image size
is limited to only 2\,$\times$\,2\,pixels due to the minimum size of one 
resolution element of 1.0\,mas. 
 
Model fitting will often be a better choice than imaging, especially
if the feasible size of the image is limited to a few pixels. 
%In addition, a model analysis can be used to determine whether or not
%a later image reconstruction will be reliable. 
The use of phase information, i.e measurements of triple products, 
is mandatory for asymmetric objects.
Recent data from the COAST interferometer on Betelgeuse, a star 
exhibiting surface spots, were in fact analyzed by model 
fits (Young et al. 2000). A model example of the influence of a 
stellar surface spot on visibility amplitude, triple amplitude, 
and closure phase data of a cool giant can be found in 
Wittkowski et al. (2001).
\section{VLTI Simulations}
It was discussed in Sect.~2 that single giants are good candidates
for interferometric imaging. As a result, the slowly rotating 
G9 giant $\Theta^1$\,Tau was chosen from Tab.~\ref{tab:largediam} 
as a model star for VLTI simulations. With a measured diameter 
of 2.6\,mas (White \& Feierman 1987) -- which is in good agreement with 
diameter estimates -- it is one of the apparently largest stars for 
which surface features by magnetic activity have been predicted. 
It is located at a declination of +16$^\circ$, which is easily 
reachable from the VLTI at Cerro Paranal (latitude -24$^\circ$ 40'). 
Its effective temperature was estimated to be 5000\,K (Choi et al. 1995). 
A surface gravity of $\log g=3$ is assumed based on empirical calibrations.

While surface activity is predicted for $\Theta^1$\,Tau, the characteristics
of its surface features are unknown. As a solution, typical spot 
parameters were assumed for our model star.
They were taken from Doppler imaging results of a similar star, 
the effectively single G8 giant CM\,Cam (see Sect.~2). The two stars differ 
in their rotational velocity (period of CM\,Cam: 16 days, 
period of $\Theta^1$\,Tau: 140 days), which may lead to a different level 
of expected surface activity. Presently, CM\,Cam as a single giant 
of about the same spectral type is the best match of a giant for which 
information on spot parameters exists.
Our model describes a CM\,Cam like star at a for VLTI
accessible location on the sky with a more convenient distance, 
i.e. angular diameter.

Cool spots with an average temperature difference with respect to the 
unspotted photosphere of $\Delta T=500\pm 300$\,K have been found on the 
surface of CM\,Cam, as well as a weaker equatorial belt of surface features
(Strassmeier et al. 1998).
The latter are
rather complex features than well defined spots, and hence difficult
structures for interferometric imaging. The 
biggest of the polar spots has been used for our model. It shows a temperature 
difference of $\Delta T=500\pm 300$\,K, and was found at a latitude 
of +65$^\circ$, with a radius of 15.3$^\circ$ (Strassmeier et al. 1998).
\paragraph{Model}
\begin{table}
\caption{Model assumptions. Some parameters are redundant and given 
for reasons of clarity only. The parameters which are directly used
for our model are marked by italic font.}
\begin{tabular}{ll}
Parameter & Value \\ \hline
Observing wavelength          & 1.0\,$\mu$m\\
Declination                   & +16$^\circ$ \\
Stellar rotation period       & 140 days / 16 days\\ 
Inclination of rotation axis  & 60$^\circ$ \\
{\it Stellar diameter}        & 2.6\,mas \\
Stellar effective temperature & 5000\,K \\
Stellar surface gravity       & $\log g=3$ \\
Metallicity                   & Solar \\
{\it Limb darkening parameter $\alpha$ for $I=\mu^\alpha$}   & 0.33 \\
%Pole position & 1.13 mas \\
{\it Spot diameter}           & 0.5 mas \\
{\it Spot separation from the stellar center} & 0.7 mas \\
{\it Position angle of the spot}  & 90$^\circ$  \\
Spot temperature              & 4500\,K \\
Ratio of spot flux to that at unspotted area & 0.71 \\
{\it Rel. offset intensity of the spot} & -0.01\\
\end{tabular}
\label{tab:modelpar}
\end{table} 
\begin{figure}
\centering
\resizebox{0.36\hsize}{!}{\includegraphics{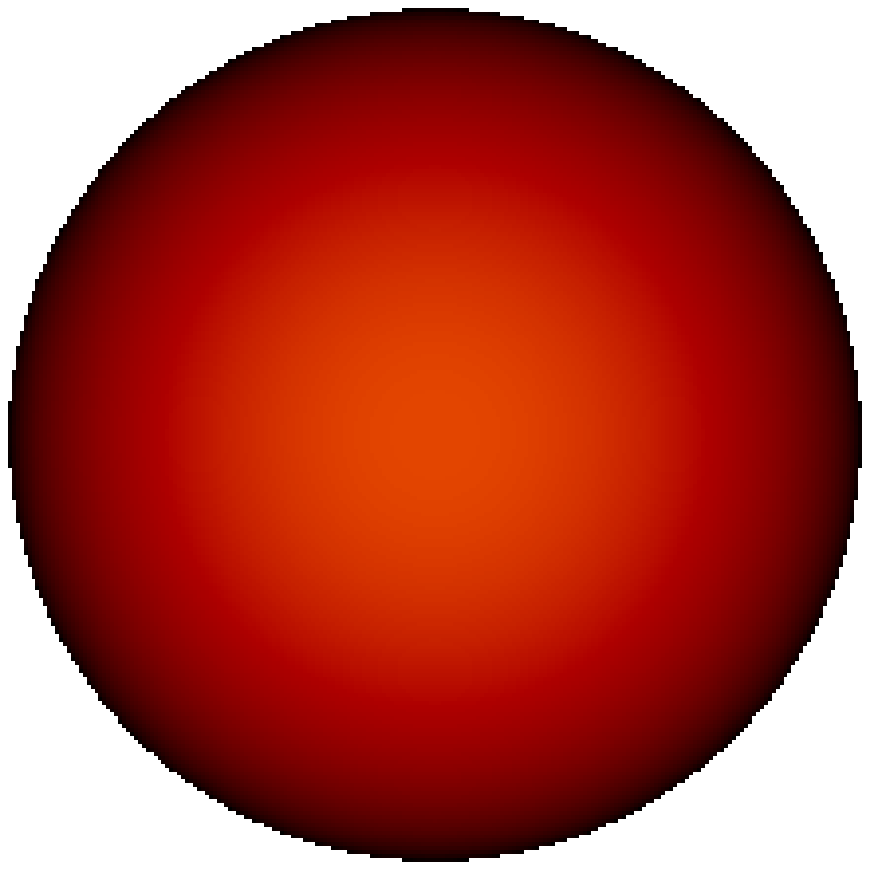}}
\resizebox{0.36\hsize}{!}{\includegraphics{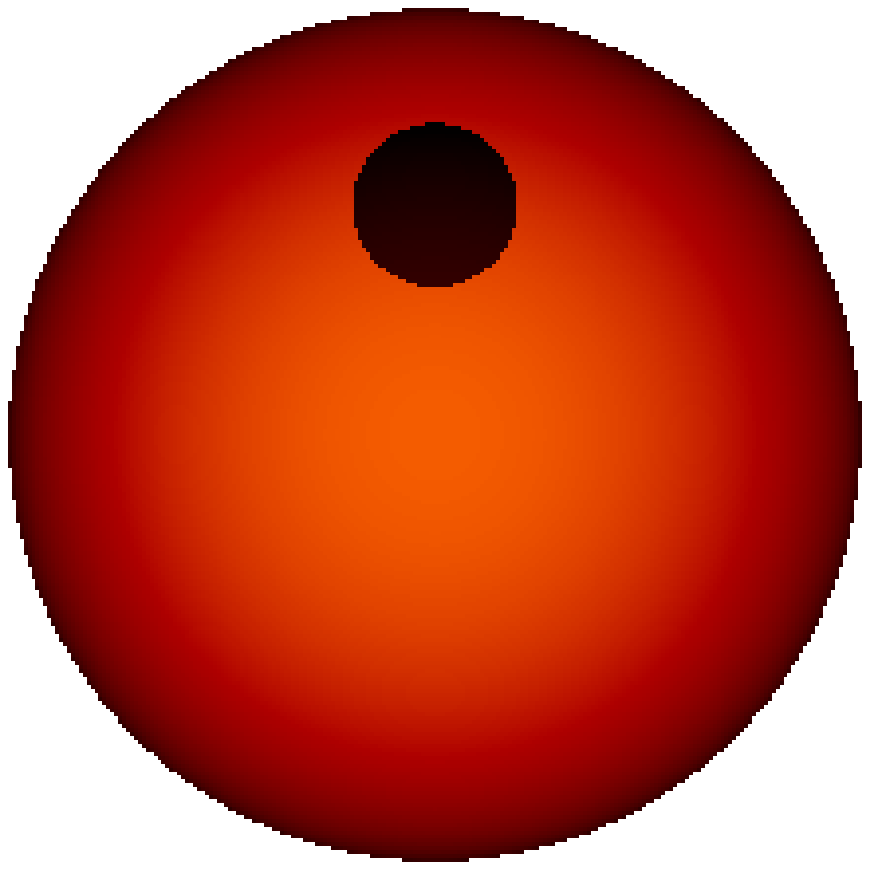}}

\resizebox{0.36\hsize}{!}{\includegraphics{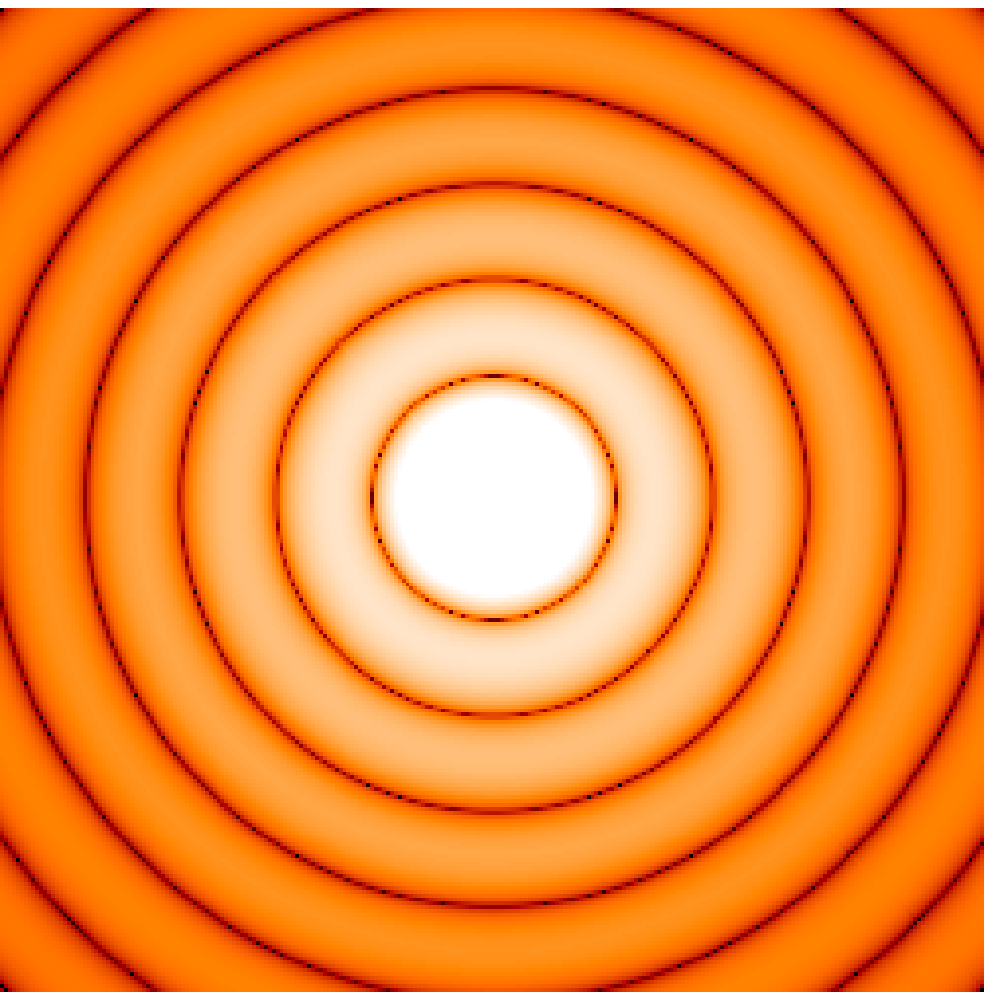}}
\resizebox{0.36\hsize}{!}{\includegraphics{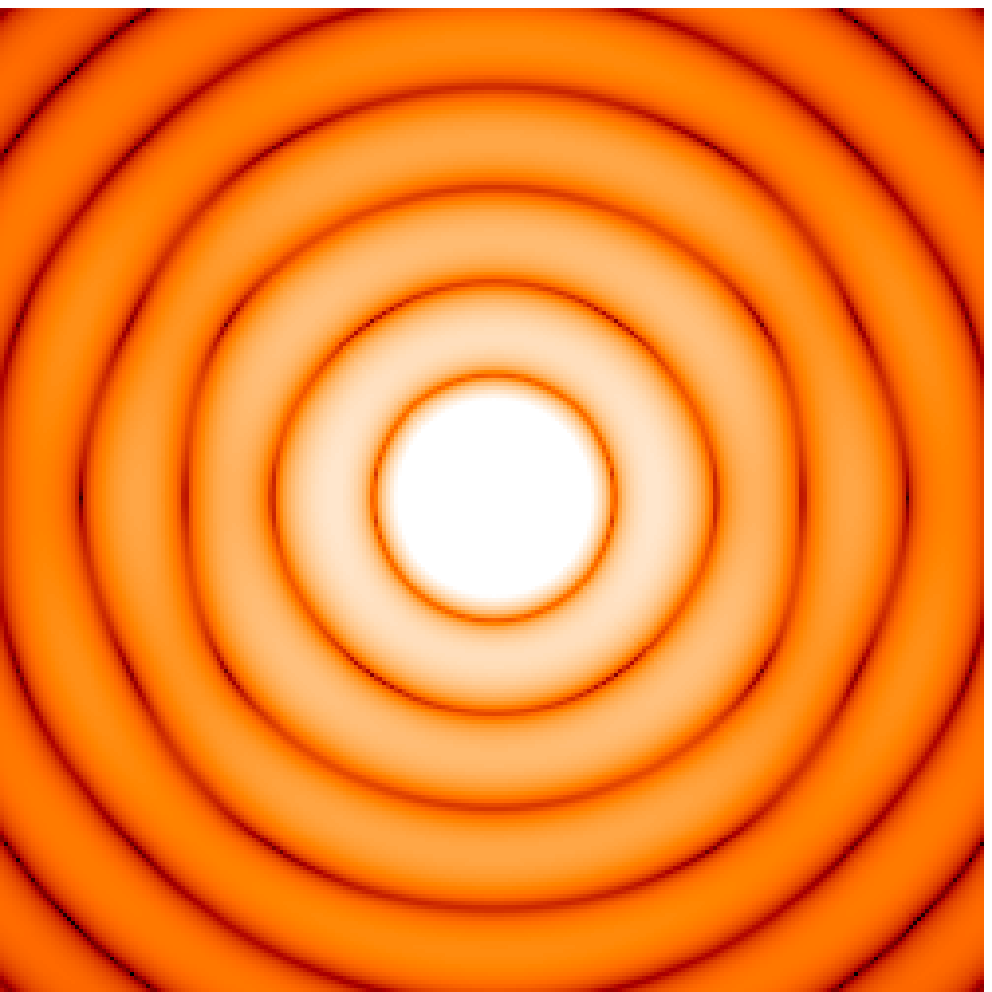}}

\resizebox{0.36\hsize}{!}{\includegraphics{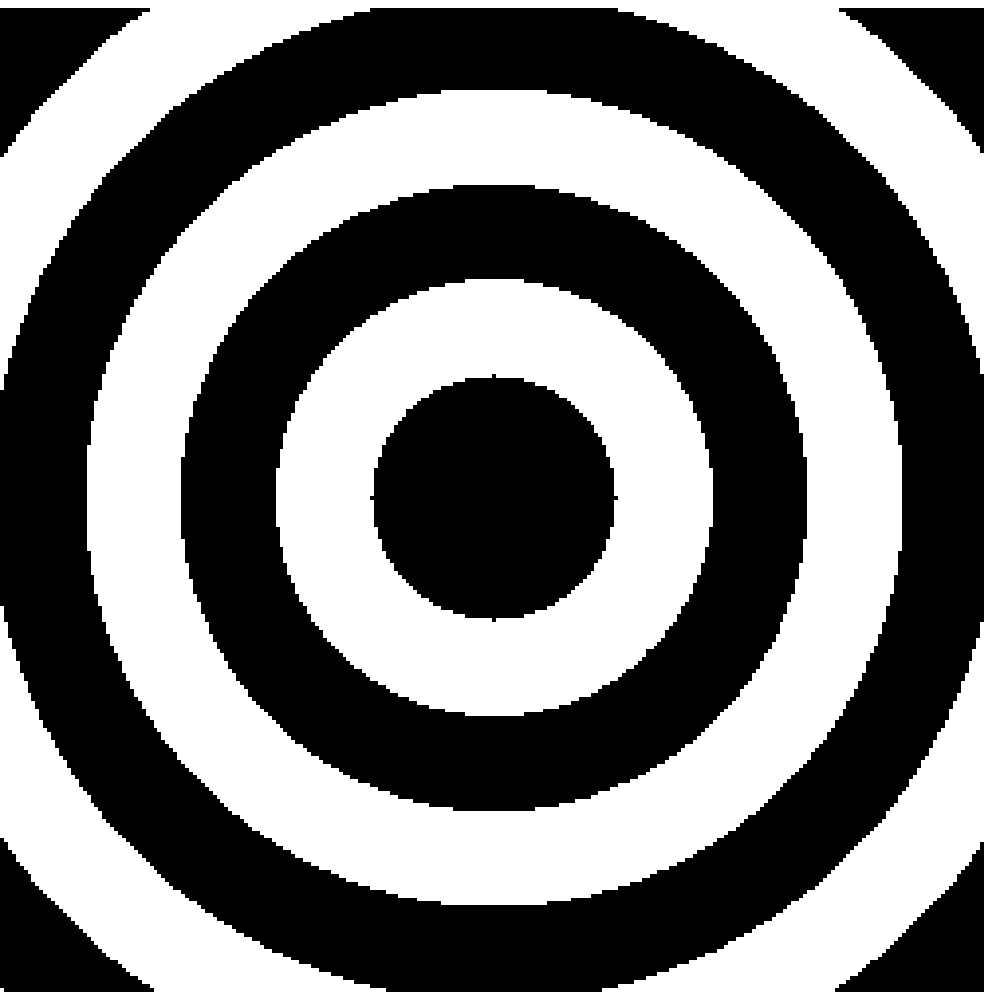}}
\resizebox{0.36\hsize}{!}{\includegraphics{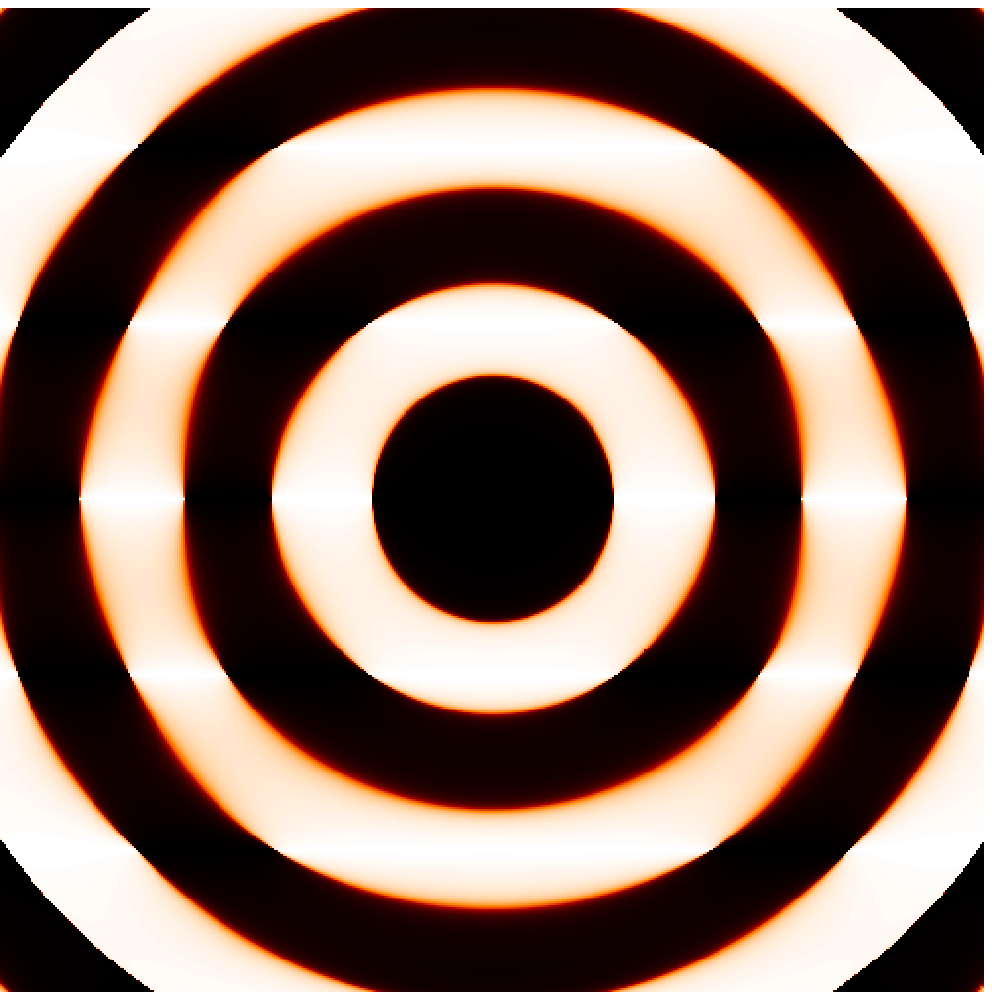}}
\caption{Illustration of our model star. Shown are the image (top), 
the visibility amplitude (middle), and the visibility phase
(bottom) of our model star (right), compared to the same star without
a surface spot (left). Model parameters were used as given in 
Tab.~\protect\ref{tab:modelpar}. The $\times$-symbol on the image of the 
model star denotes the position of the stellar pole. Visibility amplitude 
values $\in [0,0.2]$ are shown in logarithmic scale. The spatial frequency 
range for the visibility amplitudes and phases is $[-2000,2000]$ 
cycles/arcsec, corresponding to an angular resolution of up to 0.5\,mas, 
i.e. up to a projected baseline of 400\,m at a wavelength of 1\,$\mu$m. 
The VLTI with a maximum baseline of $\sim$\,200\,m operating at a wavelength
of 1\,$\mu$m can use half of the range shown here. As another example,
the NPOI with a maximum baseline of 437\,m operating at a wavelength of
about 750\,nm can use $\sim$ 1.5 times the shown range for bright stars. 
Horizontal and vertical cuts through the different panels are shown 
in Fig.~\protect\ref{fig:cuts}.
\label{fig:modelstar}}
\end{figure}
\begin{figure}
\centering
\resizebox{0.49\hsize}{!}{\includegraphics{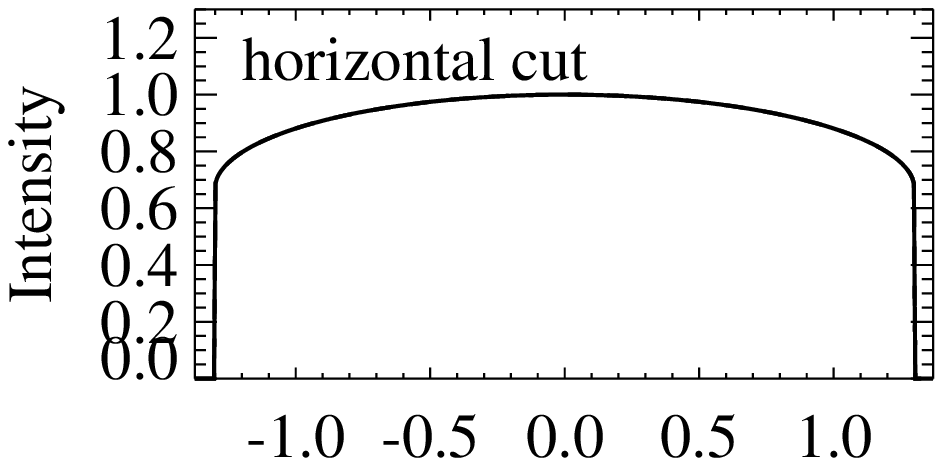}}
\resizebox{0.49\hsize}{!}{\includegraphics{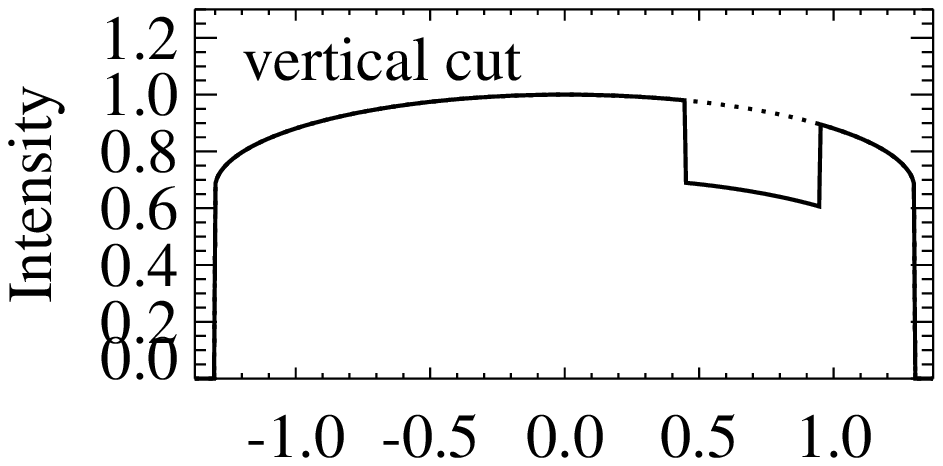}}

\vspace*{0.2cm}%
\resizebox{0.49\hsize}{!}{\includegraphics{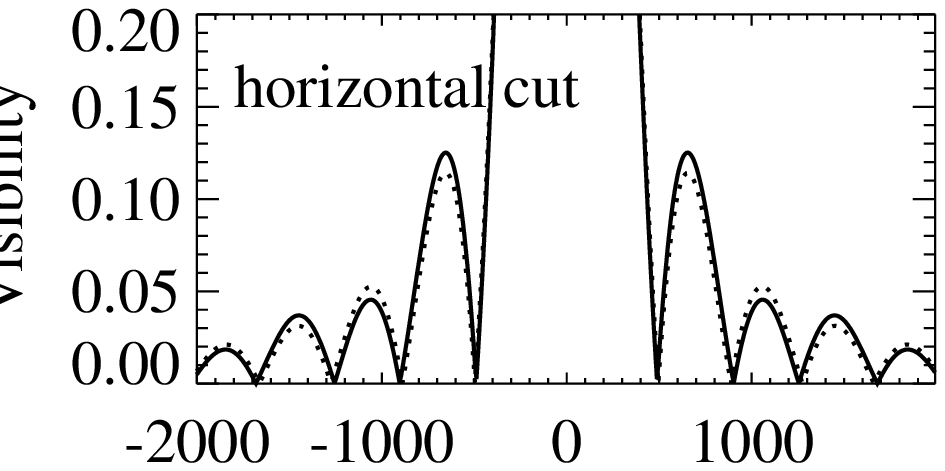}}
\resizebox{0.49\hsize}{!}{\includegraphics{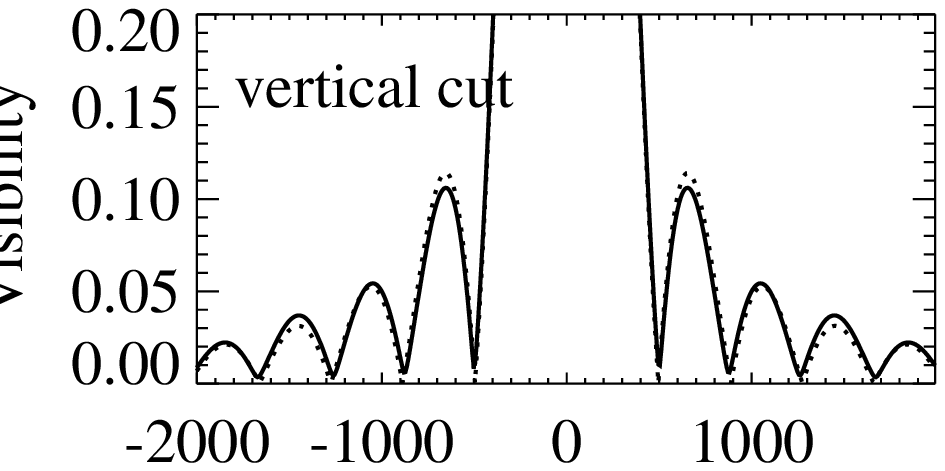}}

\vspace*{0.2cm}%
\resizebox{0.49\hsize}{!}{\includegraphics{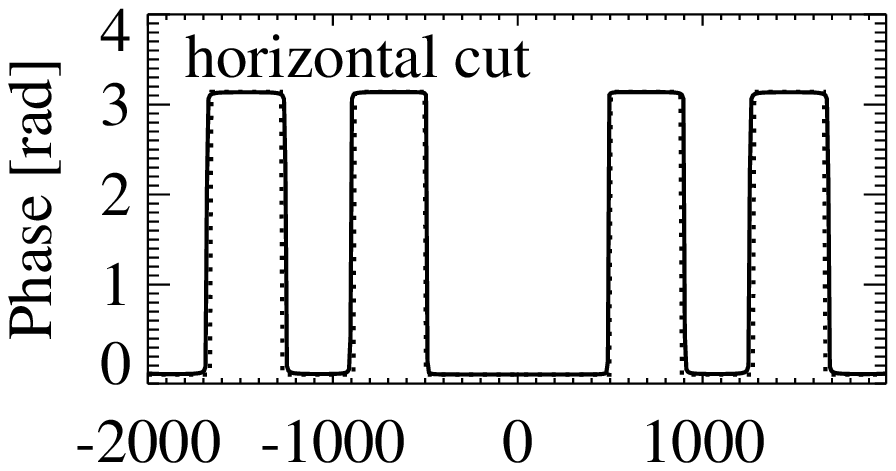}}
\resizebox{0.49\hsize}{!}{\includegraphics{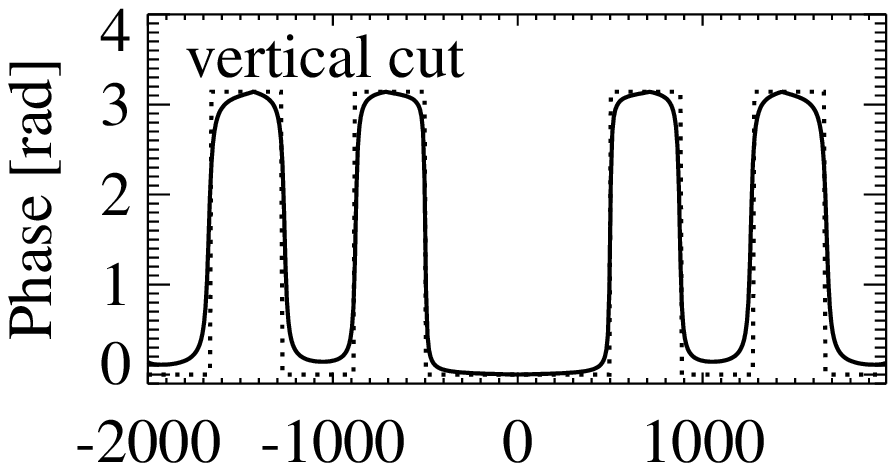}}
\vspace*{0.2cm}%
\caption{Horizontal (left) and vertical (right) cuts through the
different panels of Fig.~\protect\ref{fig:modelstar}, i.e. through 
the model images (top), visibility amplitudes (middle), and 
visibility phases (bottom). The solid lines denote our used model star, 
and the dotted lines the same star without a surface spot.
\label{fig:cuts}}
\end{figure}
A model star is considered with parameters based on the characteristics 
of $\Theta^1$\,Tau for the star parameters and CM\,Cam for the spot
parameters. The model parameters are listed in 
Table \ref{tab:modelpar}. The stellar intensity function $I$ is
described by a limb-darkened disk $I=\mu^\alpha$ with 
$\mu=\cos\theta$ being the cosine of the angle between the line of 
sight and the normal of the surface element of the star, and 
$\alpha$ a positive real number (see Hestroffer 1997).
The parameter $\alpha$ was determined so that the integral of the
normalized intensity function is the same as that of the tabulated intensity
values of the corresponding Kurucz model atmosphere
(Kurucz 1993). 
%The observed spot parameters were converted
%into an angular size of the spot, a separation of the spot center 
%from the star center, and a position angle on the stellar surface.
Figure~\ref{fig:modelstar} shows an image of our model star, as well
as the visibility amplitude and visibility phase of this image, compared
to an unspotted star. Figure~\ref{fig:cuts} shows horizontal and vertical
cuts through the different panels in Fig.~\ref{fig:modelstar}.
\paragraph{Accuracy of the AMBER instrument}
The anticipated accuracy of the AMBER instrument is given as follows. The 
numbers in brackets denote the goals.
\begin{tabular}{ll}
Contrast stability & 10$^{-2}$ in 5 min (10$^{-3}$). \\
%                   & 10$^{-3}$ in 5 min as goal.\\
Visibility accuracy & 1\% at 3\,$\sigma$ (0.01\% at 1\,$\sigma$). \\
%                    & 0.01\% at 1 $\sigma$ as goal.\\
Differential phase stability & 10$^{-2}$ rd in 1 min (10$^{-4}$\,rd). \\
%                             & 10$^{-4}$ rd in 1 min as goal.
\end{tabular}
The horizontal cuts through the visibility amplitudes
in Fig.~\ref{fig:cuts} show a relative difference between the two
models, i.e. the spotted star and the unspotted star, of $\sim$7\% 
at the maximum of the second lobe. This difference is significant 
with an accuracy of 1\% at the 3\,$\sigma$ level 
as given for the AMBER instrument. Additional calibration errors
are expected to be $\sim$\,1\%, based on the anticipated
contrast stability. As a result, it can be expected that the
given differences can well be detected. However, similar differences 
at this point of the visibility function are for example expected 
for a stellar disk with different limb-darkening parameter. 
Hence, measurements at different evenly distributed points of the $uv$-plane
have to be performed in order to unambiguously detect a 
surface spot (see discussion in Sect.~3). For instance, a stellar disk
without a surface feature will not lead to the flip of the two model
visibility amplitudes between the second and fourth lobe.
In addition, deviations of the visibility phase from values zero and $\pi$
are an unambiguous signature of asymmetric structures and, hence, an 
essential indicator for interferometric measurements of stellar surface
structures. 
The deviations of our model star's visibility 
phase from values zero and $\pi$ as shown in Fig.~\ref{fig:cuts} are 
well above the expected AMBER phase stability of $\sim$\,0.01\,rd.

\begin{figure}
\resizebox{0.85\hsize}{!}
{\includegraphics{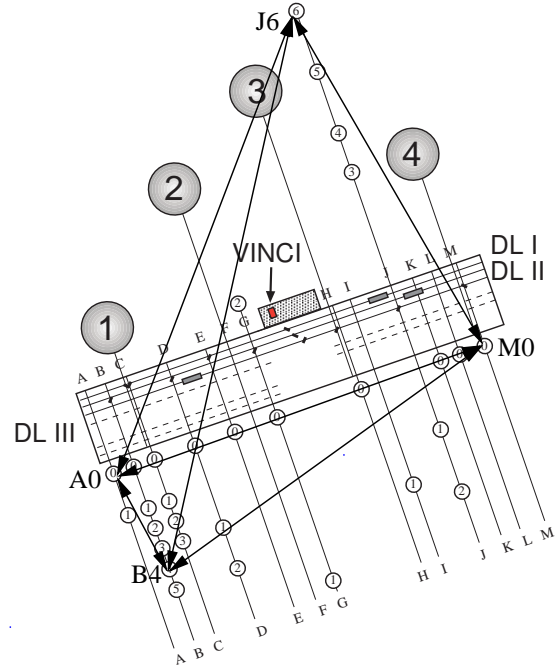}}
%{\includegraphics{vlt-area.eps}}
\caption{General layout of the VLTI. Shown are the positions of the
4 UTs, the 30 AT stations, the system of light ducts, the delay line
tunnel with the currently existing three delay lines (three more
are expected to be installed by mid 2003), and the interferometric lab with 
the currently used commissioning instrument VINCI. The arrows
indicate the baselines which were used in our simulation, i.e. those 
between stations B4, J6, A0, and M0. The baseline length between 
stations B4 and J6 is 195.31\,m, that between stations A0 and M0 is 144.0\,m.
North is at the top.}
\label{fig:layout}
\end{figure}

\begin{figure}
\resizebox{0.95\hsize}{!}
{\includegraphics{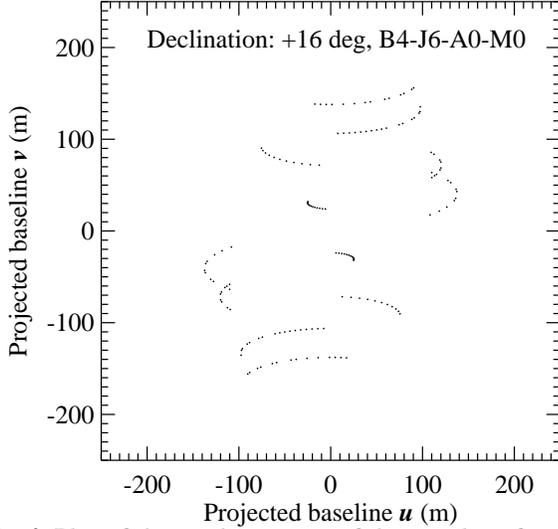}}
\vspace*{-1cm}%
\caption{Plot of the used coverage of the $uv$-plane for an object
at a declination of $+16^\circ$ ($\Theta^1$\,Tau). The size of each 
circle represents the size of a 1.8\,m 
diameter VLTI Auxiliary Telescope. Four different telescopes are used 
at positions B4, J6, A0, M0. One out of four possible triple configurations
is measured every 15 min. Tab.~\protect\ref{tab:triples} shows which
triple configuration was used at which time (based on observing
restrictions of some baselines owing to limited delay line lengths).}
\label{fig:uv}
\end{figure}

For the observables, namely the squared visibility amplitude,
the triple amplitude, and the closure phase, conservative accuracies
including calibration errors of 4\%, 4\%, and 0.3 rd, respectively, 
are assumed in the following.
\paragraph{Choice of baselines and $uv$ coverage}
The rotation period of $\Theta^1$\,Tau of 140 days would allow observations
during two consecutive nights with a smearing effect of less than 5$^\circ$.
This makes it possible to reconfigure the array from one night to the
other and to combine the data. However, since our model is also intended
to describe a CM\,Cam type star at the location of $\Theta^1$\,Tau,
the CM\,Cam rotation period of only 16 days limits our combinable 
observing time to about 6 hours during only one night, with a smearing 
effect of less than about 5$^\circ$\footnote{In case that
the data are intended to be analyzed by model fits only, the rotation
of the star can be modeled as an additional parameter, and the data
from many nights can be combined. Then, more complete coverages of the
$uv$-plane can be obtained.}. Due to the use of data taken during only one 
night, a reconfiguration of the array is not possible.
In order to obtain a good coverage of the $uv$-plane despite of this,
four telescopes are used for our simulation. The beams of any 
three of them can be fed to the AMBER instrument at any time.
Four different triple configurations can then be used, of
which three are independent.
Not all baselines can be used at any time due to a limited range of the
delay lines. Observations are simulated for 6 hours,
3 hours before meridian until 3 hours after meridian, with one measurement
every 15 minutes on the next available triple configuration.
Stations B4, J6, A0, and M0 were selected in order to achieve a 
maximum angular resolution at all position angles. Figure~\ref{fig:layout}
shows a schematic view of the VLTI layout indicating the used array stations
and baselines. Table~\ref{tab:triples} 
shows when each of the four possible triple configurations is used.
Figure \ref{fig:uv} shows the obtained coverage of the $uv$-plane.
The selection of the telescope stations, as well as the calculation of the
resulting coverage of the $uv$-plane considering the different restrictions
was performed using observation preparation tools as described 
by Sch\"oller (2000). 
\begin{table}
\caption{The four different used triple configurations and the times
when each of them is used, in hours relative to meridian.}
\label{tab:triples}
\begin{tabular}{l|rrrrrr}
\hline
B4 J6 A0 & -3.00 & -2.75 & -2.50 & -2.25 & -1.75& -0.75 \\
         & 0.25  & 1.25  & 2.25  &       &      &       \\ \hline
B4 J6 M0 &-1.25  &-0.5   & 0.5   & 1.5   & 2.5  &       \\ \hline
B4 A0 M0 &-0.25  & 0.75  & 1.75  & 2.75  &      &       \\ \hline
J6 A0 M0 &-2.00  & -1.50 & -1.00 & 0.00  & 1.00 & 2.00  \\ 
         & 3.00  &       &       &       &      &       \\ \hline
\end{tabular}
\end{table}
\paragraph{Model results}
\begin{table}
\label{tab:bestfit}
\caption{Reduced $\chi^2_n$ value, stellar diameter, and 
limb-darkening parameter $\alpha$ of that symmetric stellar disk without 
surface feature which fits best our simulated model star.
The model star with parameters as given in Tab.~\protect\ref{tab:modelpar} 
(diameter 2.6\,mas) was used, as well as models for the same star at
twice the distance (diameter 1.3\,mas) and half the 
distance (diameter 5.2\,mas).}
\begin{tabular}{lrrr}
Model star diameter (mas)& 2.60 & 1.30  & 5.20 \\
\hline
Stellar diameter (mas)   & 2.68 &  1.34 & 5.42     \\
Limb-darkening parameter & 0.52 &  0.56 & 0.62     \\
Reduced $\chi^2$         & 16.0 &  0.1  & 186.7    \\
\end{tabular}
\end{table}
\begin{table}
\caption{Accuracy ranges of our model parameters for a 
5 $\sigma$ detection. The values are derived by varying
each of the parameters and calculating the reduced
$\chi^2_n$ value. The stellar diameter was adjusted to 
its best fitting value for every variation while all other
parameters were kept constant.}
\begin{tabular}{lrr}
Parameter & Model value & Accuracy range\\ \hline
Limb darkening par. & 0.33 & [0.27..0.41] \\
Spot diameter [mas]  & 0.5 & ]0..0.7] \\
Spot separation [mas] & 0.70 & [0.67..0.72] \\
Spot P.A. [degree]    & 90 & [87..101] \\
Rel. offset intensity [\%] & -1.0 & [-0.8..-1.3] \\ 
\end{tabular}
\label{tab:accuracy}
\end{table}
\begin{figure}
\resizebox{0.95\hsize}{!}
{\includegraphics{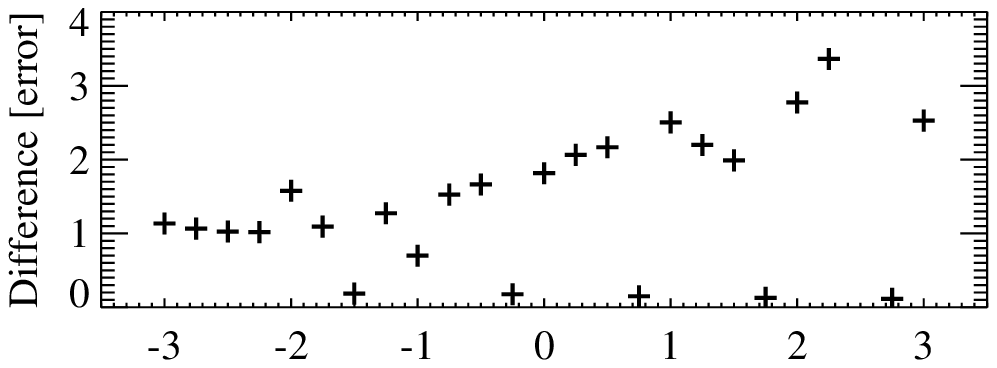}}

\resizebox{0.95\hsize}{!}
{\includegraphics{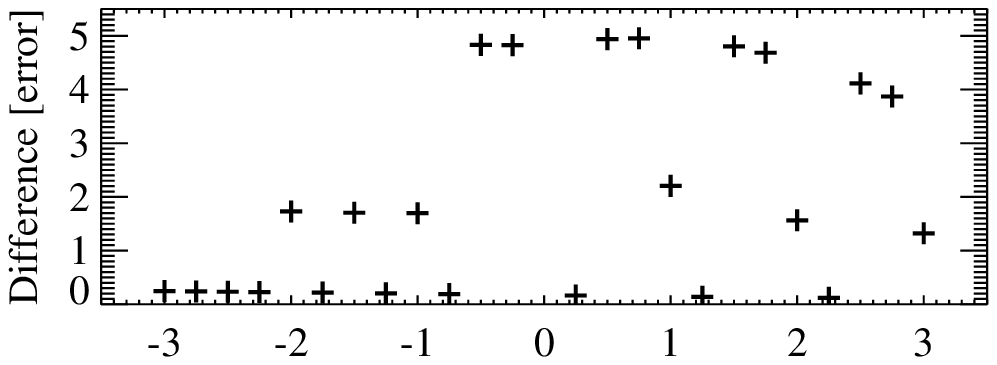}}

\resizebox{0.95\hsize}{!}
{\includegraphics{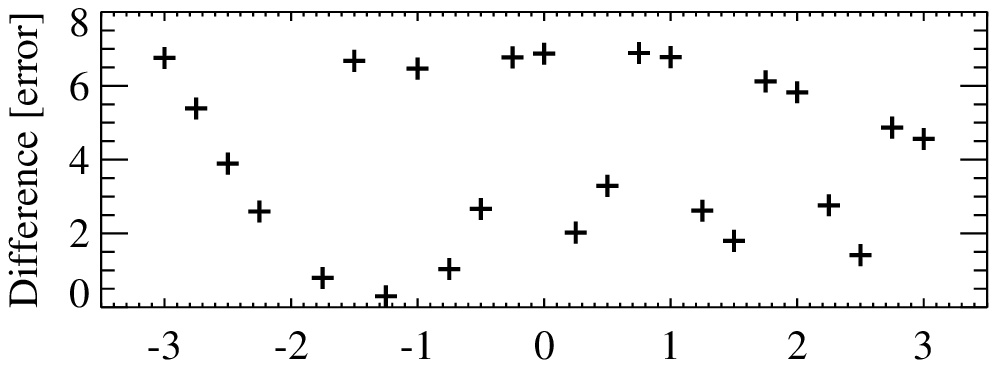}}

\resizebox{0.95\hsize}{!}
{\includegraphics{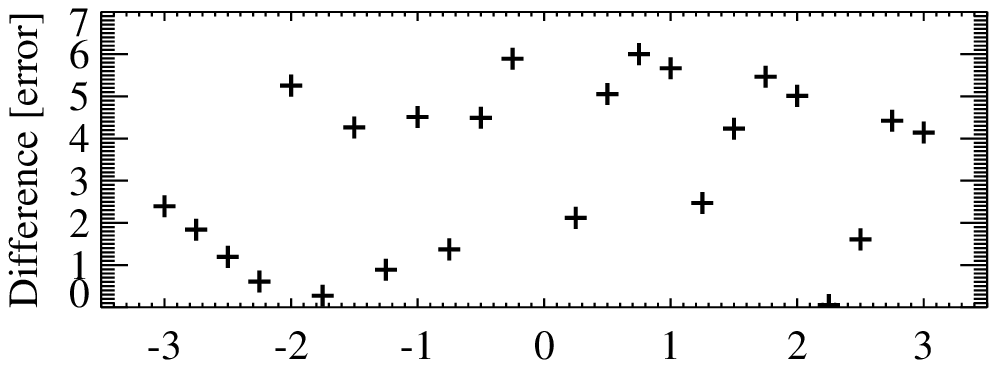}}

\resizebox{0.95\hsize}{!}
{\includegraphics{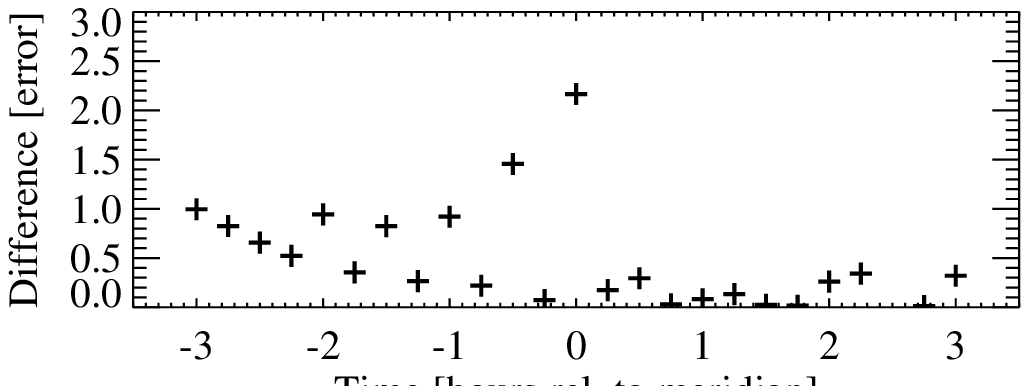}}
\caption{Difference of our three model squared visibility amplitudes,
triple amplitudes, and closure phase values (from top to bottom) to 
those of the best fitting symmetric stellar disk without a surface feature, 
in units of the size of the assumed error bar. The x-axis shows the
time of the observation relative to meridian. Observations are simulated
according to the triple configurations shown 
in Tab.~\protect\ref{tab:triples}.}
\label{fig:diff}
\end{figure}
The three squared visibility amplitudes, the triple amplitude, and
the closure phase of our model star were calculated for each of the 
25 observations which are listed in Tab.~\ref{tab:triples}.
The total number of independent data points of our simulation is 100.  
Errors were calculated according to the adopted precisions of the AMBER
squared visibility amplitudes, triple amplitudes, and closure phases.

In order to determine whether the surface spot on our model star can be 
detected, the reduced $\chi^2_n$ value of the best fitting symmetric 
stellar disk without a surface spot to our spotted model star was calculated. 
The stellar diameter was treated as the only free parameter. Fits were 
performed for a range of assumed limb-darkening parameters. The minimum 
reduced $\chi^2_n$ value that was found during this procedure has been 
derived. In addition, these calculations were performed for the same 
model star shifted to twice its distance (diameter 1.3\,mas) and half its 
distance (diameter 5.2\,mas). 
Table~\ref{tab:bestfit} shows the resulting values. The asymmetric feature 
on our model star can be detected with a high significance ($\chi^2_n=16$). 
The same star with an apparent diameter of 1.3\,mas can not be distinguished
from a symmetric stellar disk ($\chi^2_n=0.1$), while the significance of
the asymmetric feature on the 5.2\,mas star is considerably 
higher ($\chi^2_n=187$). Figure~\ref{fig:diff} shows the differences of
our simulated model star squared visibility amplitude, triple amplitude,
and closure phase values to that of the best fitting symmetric
stellar disk, in units of the assumed sizes of the error bars.

In order to estimate the accuracy of fitted model parameters,
each of the parameters of our model star was varied 
until the reduced  $\chi^2_n$ value reached a value of 5. The stellar
diameter was adjusted to its best fitting value after every variation
while all other parameters were kept constant. Table~\ref{tab:accuracy} 
shows the resulting accuracy ranges in comparison to our model 
as given in Tab.~\ref{tab:modelpar}. Based on our assumptions, 
all model parameters can be derived with reasonable accuracy. The spot 
itself (0.5\,mas diameter) is unresolved, with our conservative accuracy 
assumptions and a required 5\,$\sigma$ detection. Thus, our model star 
can not be distinguished from a star with a smaller spot if the spot's 
relative offset intensity remains in the range [-0.8\%..-1.3\%]. In other
words, a smaller spot with a larger temperature difference relative to 
the unspotted photosphere can not be distinguished from a larger spot 
(up to 0.7\,mas) with a smaller temperature difference.  
If, however, our model star is shifted to half its distance (stellar diameter
5.2\,mas, spot diameter 1\,mas), the spot diameter would be constrained
to the range [0.85\,mas..1.15\,mas], and together with the constrains for
the spot's offset intensity, the spot temperature could be derived.
\section{Summary and discussion}
We have investigated the feasibility to image surface spots on
magnetically active stars with the VLT interferometer. We used
realistic model assumptions based on a Doppler image of the giant
star CM\,Cam. Observations were considered which make use of the 
first generation VLTI instrument AMBER. All observing restrictions
were taken into account, as for example the limited lengths of 
the delay lines. We used conservative assumptions for the 
accuracy and calibration uncertainty of the AMBER instrument. 
We have discussed that magnetically active single giants are the 
most promising candidate stars for interferometric 
imaging. We show by simulations of observations which are
performed during only one night, that the surface feature on our 
model giant can be detected at a high significance level and estimate 
the accuracy of fitted spot parameters to be reasonably good. 
The combination of data taken during one night limits the rotational
period of the star to about 15-20 days. For faster rotating
stars, the rotation period could be modeled as an additional parameter. 
Our analysis shows as well that the angular diameter of our model 
star of 2.6\,mas is at the lower limit for feasible interferometric 
studies of stellar surfaces using the VLTI with AMBER. We find, on the 
other hand, that the majority of stars for which the existence of 
surface features caused by magnetic activity is known or strongly 
suggested, have angular diameters smaller than 2\,mas and, in addition, 
are located in the northern hemisphere. It would be desirable to initiate 
spectroscopic and photometric variability studies in the southern hemisphere, 
focusing on apparently large giants, as a preparation and source of 
complimentary information for VLTI measurements of stellar surface features. 
The VLTI, other ground based 
interferometers, and finally large space based UV interferometers promise 
to give us exciting new insights into magnetic and hydrodynamic activity 
of stars, with first results to be expected very soon.
\acknowledgements This research has made use of the SIMBAD database,
operated at CDS, Strasbourg, France.

\begin{appendix}
\section{List of estimated angular diameters}
\label{sec:diam}
Angular diameters were estimated for stars with known or expected
surface structures in order to find suitable stars for interferometric 
observations of surface inhomogeneities. 
As a primary source, the ''Summary of Doppler images of late-type stars
(www.aip.de/groups/activity/DI/summary) was used. In addition, a list
of stars exhibiting photometric variations (Strassmeier et al. 1999)
was used. For the latter stars, periods and radial velocities are
known, but indirect surface images are not available. Some stars are
listed in both of these sources. Furthermore, a list of slowly
rotating single giants exhibiting Ca\,II variability (Choi et al. 1995)
was used. These stars are expected to show surface structures but
additional information on spot parameters is not available.

The following methods were employed in order to estimate stellar angular
diameters:
\begin{enumerate}
\item The absolute diameter was calculated based on the effective
  temperature and bolometric luminosity. Tabulated values in 
  Schmidt-Kaler (1982), based on the spectral type, were used for 
  these quantities. The Hipparcos parallax was used to
  derive the angular diameter.  
\item The empirical calibration for the angular diameter by Dyck et
  al. (1996) based on the spectral type and the 
  K magnitude (Gezari et al. 1999) was used
  for giants.  
  In case that
  the measured K magnitude includes a companion, the
  resulting error on the diameter is of the order of one.
\item The photometric period and the rotation velocity $v \sin i$ 
  were used to derive the absolute radius of the star.  This
  radius was converted into an angular diameter using the Hipparcos
  parallax. If the angle of inclination of the spot is not known, this
  estimate is a lower limit. In case that Doppler images
  exist, the angle of inclination given and the
  angular diameter was computed.
\item The CHARM (catalog of high angular resolution
  measurements, Richichi \& Percheron 2002) was searched for
  measurements of angular diameters.
\end{enumerate}
The errors on the estimated diameters can amount to up to a factor 
of about 2. Within these errors, the different
estimates are usually consistent. Cases with larger deviations include
the RS\,CVn variables CF Tuc, IN Vir, and II Peg, as well as the T
Tauri stars V987 Tau, SU Aur, HT Lup, and V824 Tau. 
\begin{table*}[bt]
\caption{\label{tab:diamest}Estimated diameters for stars with 
known surface features. The sources are (Ref 2) "Summary of 
Doppler images of late type stars", 
www.aip.de/groups/activity/DI/summary (version from Dec. 18, 2001) and 
(Ref 1) the list in ''Starspot photometry with robotic telescopes'', 
Strassmeier et al. (1999).
Listed are the star's name, its HD number, the type of variability as given 
in the references, the parallax $\pi$ taken from Perryman \& ESA (1997), 
the K magnitude from Gezari et al. (1999), if available, the spectral type,
the rotation period P, the rotational velocity $v \sin i$, all as given in the
references, the three different estimates of the angular diameters 
($\theta_1$, $\theta_2$, $\theta_3$), and, finally, a measured value 
for the angular diameter, if available. The three different methods
for the diameter estimates are explained in the text. 
The measured diameters are limb-darkened diameters, averaged over different
measurements if more than one measured value was available. The references 
for the measured diameters are as follows. $\zeta$ And: Cohen et al. (1999), 
Hutter et al. (1989); V833 Tau: Chen \& Simon (1997); 
SU Aur: Akeson et al. (2000, UD diameter in K-band); $\epsilon$ Aur: 
Nordgren et al. (2001); $\sigma$ Gem: Cohen et al. (1999) and 
Nordgren et al. (1999); 31 Com: Bell \& Gustaffson (1989).
}
\begin{tabular}{lrlrrlrrrrrrrrrrr}
\hline
Name  & HD  & Type  &$\pi$& SpT &  K & P &  P &$v \sin i$ & $v \sin i$ & $i$ & $\theta_1$ & $\theta_2$  & 
$\theta_3$ & $\theta_m$  \\
      &     &       &     &     &    & \small Ref 1&\small Ref 2 &\small Ref 1     &\small Ref 2      &     &            &  & & \\
      &     &       & mas &   & mag & d &  d &km/s   & km/s  & deg & mas  & mas & mas & mas \\
\hline
V640 Cas     &    123 &  SB       &  49.3 & G8V   &      &        &       &  4.7 &      &    & 0.4 &     &            & \\
LN Peg       &        &  RS CVn   &  24.7 & G8V   & 6.4 &   1.85 &       & 23   &      &    & 0.2 &     & $\ge$ 0.2  & \\
$\zeta$ And  &   4502 &  RS CVN   &  18.0 & K1III & 1.4 &        &17.64  &      &  41  &    & 2.1 & 2.6 & $\ge$ 2.4  & 2.8 \\
CF Tuc       &   5303 &  RS CVn   &  11.6 & K4IV  & 5.3 &        &  2.8  &      &  35  & 64 & 0.9 & 0.5 & 0.2        & \\
AE Phe A     &   9528 &  W UMa    &  20.5 & G0V   &      &        &  0.36 &      & 145  & 88 & 0.2 &     & 0.2        & \\
AE Phe B     &   9528 &  W UMa    &  20.5 & F8V   &      &        &  0.36 &      &  95  & 88 & 0.2 &     & 0.1        & \\
XX Tri       &  12545 &  RS CVn   &   5.1 & K0III &      &  23.87 & 24    & 18.2 &  21  & 60 & 0.5 &     & 0.5        & \\
VY Ari       &  17433 &  RS CVn   &  22.7 & K3IV  &      &  16.23 &       & 10   &      &    & 1.4 &     & $\ge$ 0.7  & \\
UX Ari       &  21242 &  RS CVn   &  19.9 & K0IV  & 3.8 &        &  6.44 &      &  39  & 60 & 0.9 & 0.8 & 1.1        & \\
V711 Tau     &  22468 &  RS CVn   &  34.5 & K1IV  & 3.3 &   2.84 &  2.84 & 40   &  41  & 35 & 1.6 & 1.1 & 1.3        & \\  
V471 Tau     &        &  BY Dra   &  21.4 & K0V   &      &        &  0.52 &      &  91  & 90 & 0.2 &     & 0.2        & \\
EI Eri       &  26337 &  RS CVn   &  17.8 & G5IV  &      &   1.91 &  1.95 & 50   &  51  & 46 & 0.6 &     & 0.4        & \\
YY Eri A     &  26609 &  W UMa    &  18.0 & G9V   &      &        &  0.32 &      & 160  & 82 & 0.1 &     & 0.2        & \\
YY Eri B     &  26609 &  W UMa    &  18.0 & G7V   &      &        &  0.31 &      & 110  & 82 & 0.2 &     & 0.1        & \\
V410 Tau     & 283518 &  WTTS     &   7.2 & K4    & 7.5  &   1.87 &  1.87 & 77   &  77  & 70 &     &     & 0.2        & \\
EK Eri       &  27536 &  giant    &  15.8 & G8III &      & 306.9  &       &  1.5 &      &    & 1.5 &     & $\ge$ 1.3  & \\
RY Tau       & 283571 &  CTTS     &   7.5 & F8V   & 6.2  &        &       & 49   &      &    & 0.1 &     &            & \\ 
V987 Tau     & 283572 &  WTTS     &   7.8 & G5III & 6.8  &   1.53 &  1.55 & 78   &  78  & 48 & 0.6 & 0.2 & 0.2        & \\
DF Tau       & 283654 &  CTTS     &  25.7 & M0    & 6.8  &        &  8.5  &      &  25  & 60 &     &     & 1.2        & \\
V833 Tau     & 283750 &  BY Dra   &  56.0 & K5V   &      &   1.81 &       &  6.3 &      &    & 0.4 &     & $\ge$ 0.1  & $< 10$\\
SU Aur       & 282624 &  CTTS     &   6.6 & G2    & 5.8  &        &  3.09 & 66   &  64  & 80 &     &     & 0.3        & 1.9 \\
YY Men       &  32918 &  FK Com   &   3.4 & K2III &      &        &  9.55 &      &  50  & 50 & 0.5 &     & $\ge$ 0.4  & \\
$\epsilon$ Aur &31964 &           &   1.6 & F0Ia  & 1.5 &        &       &      &      &    & 1.6 & 1.8 &            & 2.2 \\
V1192 Ori    &  31993 &  giant    &   4.2 & K2III &      &  26.7  &       & 33   &      &    & 0.6 &     & $\ge$ 0.7  & \\
V390 Aur     &  33798 &  giant    &   8.9 & G8III &      &   9.69 &       & 33   &      &    & 0.8 &     & $\ge$ 0.5  & \\
AB Dor       &  36705 &  Single   &  66.9 & K0V   & 4.7 &        &  0.51 &      &  91  & 60 & 0.5 &     & 0.7        & \\
V1355 Ori    & 291095 &  RS CVn   &  23.1 & K1IV  &      &   3.87 &       & 46   &      &    & 1.1 &     & $\ge$ 0.8  & \\
V1358 Ori    &  43989 &  RS CVn   &  20.1 & G0III &      &   3    &       & 42   &      &    & 1.6 &     & $\ge$ 0.5  & \\
SV Cam       &  44982 &  RS CVn   &  11.7 & G2V   & 7.4 &        &  0.59 &      & 117  & 90 & 0.1 &     & 0.2        & \\
CM Cam       &  51066 &  giant    &   3.6 & G8III &      &  16.0  & 16    & 47   &  47  & 60 & 0.3 &     & 0.6        & \\
$\sigma$ Gem &  62044 &  RS CVn   &  26.7 & K1III & 1.7 &  19.62 & 19.4  & 27   &  27  & 60 & 3.2 & 2.3 & 3.0        & 2.4\\
TY Pyx       &  77137 &  RS CVn   &  17.9 & G5V   &      &        &  3.20 &      &  35  & 88 & 0.2 &     & 0.4        & \\
IL Hya       &  81410 &  RS CVn   &   8.4 & K1III &      &  12.67 & 12.7  & 26.5 &  26.5& 55 & 1.0 &     & 0.6        & \\
DX Leo       &  82443 &  BY Dra   &  56.4 & K0V   &      &   5.43 &       &  5   &      &    & 0.4 &     & $\ge$ 0.3  & \\
LQ Hya       &  82558 &  BY Dra   &  54.5 & K2V   &      &   1.60 &  1.61 & 28   &  27  & 60 & 0.4 &     & 0.5        & \\      
$\xi$ UMa B  &  98230 &  SB       & 137.0 & K2V   & 2.3 &        &       &  2.8 &      &    & 1.0 &     & $\ge$      & \\
DM UMa       &        &  RS CVn   &   7.2 & K0III &      &        &  7.5  &      &  26  & 40 & 0.8 &     & 0.4        & \\
HU Vir       & 106225 &  RS CVn   &   8.0 & K0III &      &  10.66 & 10.4  & 27   &  25  & 61 & 0.9 &     & 0.5        & \\
             & 111395 &  solar    &  58.2 & G5V   &      &  15.8  &       &  2.9 &      &    & 0.5 &     & $\ge$ 0.5  & \\
31 Com       & 111812 &  giant    &  10.6 & G0III & 3.3 &   6.96 &       & 57   &      &    & 0.9 & 1.0 & $\ge$ 0.8  & 0.9 \\
LW Hya       &        &  CBPN     &   7.5 & G8III &      &        &  0.76 &      &  90  & 45 & 0.7 &     & 0.1        & \\
IN Com       & 112313 &  RS CVn   &   0.8 & G5III &      &   5.90 &  5.91 & 67   &  67  & 45 & 0.1 &     & 0.1        & \\ 
37 Com       & 112989 &  giant    &   3.6 & G8III &      &        &       &  4   &      &    & 0.3 &     &            & \\
IN Vir       & 116544 &  RS CVn   &   8.8 & K2III &      &        &  8.23 &      &  24  & 60 & 1.2 &     & 0.4        & \\
FK Com       & 117555 &  FK Com   &   4.0 & G2III &      &   2.41 &  2.4  &160   & 155  & 65 & 0.3 &     & 0.3        & \\
EK Dra       & 129333 &  single   &  29.5 & G2V   &      &   2.60 &  2.6  & 17.5 &  17.3& 60 & 0.3 &     & 0.3        & \\
UV CrB       & 136901 &  RS CVn   &   3.6 & K2III &      &  18.66 &       & 42   &      &    & 0.5 &     & $\ge$ 0.5  & \\
UZ Lib       &        &  RS CVn   &   7.1 & K0III & 6.4 &   4.75 &  4.74 & 67   &  69  & 30 & 0.8 & 0.3 & 0.8        & \\
$\alpha$ CrB & 139006 &  eclips   &  43.7 & A0V   & 2.1 &        &       & 13   &      &    & 1.1 &     &            & \\
$\gamma$ CrB & 140436 &  Maia     &  22.5 & A0V   &      &   0.45 &       &100   &      &    & 0.6 &     & $\ge$ 0.2  & \\
\hline
\end{tabular}
\end{table*}
\begin{table*}[b]
\caption{Table \protect\ref{tab:diamest}, continued}
\begin{tabular}{{lrlrrlrrrrrrrrrrr}}
\hline
Name  & HD  & Type  &$\pi$& SpT &  K & P &  P &$v \sin i$ & $v \sin i$ & $i$ & $\theta_1$ & $\theta_2$  & 
$\theta_3$ & $\theta_m$  \\
      &     &       &     &     &    &\small Ref 1&\small Ref 2 &\small Ref 1     &\small Ref 2      &     &            &  & & \\
      &     &       & mas &   & mag & d &  d &km/s   & km/s  & deg & mas  & mas & mas & mas \\
\hline
HT Lup       &        &  CTTS     &   6.3 & K2III & 6.5 &        &  3.9  &      &  40  & 70 & 0.9 & 0.3 & 0.2        & \\
$\delta$ CrB & 141714 &  giant    &  19.7 & G5III & 2.7 &  57    &       &  5   &      &    & 1.5 & 1.3 & $\ge$ 1.0  & \\
V2253 Oph    & 152178 &  RS CVn   &   2.1 & K0III &      &  22.07 &       & 28.8 &      &    & 0.2 &     & $\ge$ 0.3  & \\
V824 Ara     & 155555 &  WTTS     &  31.8 & G5IV  &      &        &  1.68 &      &  35  & 53 & 1.2 &     & 0.4        & \\
V889 Her     & 171488 &  solar    &  26.9 & G0V   &      &   1.34 &       & 33   &      &    & 0.3 &     & $\ge$ 0.2  & \\
PZ Tel       & 174429 &  WTTS     &  20.1 & K0V   &      &        &  0.94 &      &  68  & 65 & 0.2 &     & 0.3        & \\
V1794 Cyg    & 199178 &  FK Com   &  10.7 & G5III & 3.2 &   3.34 &  3.32 & 67   &  71.5& 50 & 0.8 & 1.1 & 0.6        & \\
ER Vul       & 200391 &  BY Dra   &  20.1 & G0V   &      &        &  0.69 &      &  85  & 67 & 0.2 &     & 0.2        & \\
LO Peg       &        &  Single   &  39.9 & K5V   &      &        &  0.42 &      &  69  & 35 & 0.3 &     & 0.4        & \\
V2075 Cyg    & 208472 &  RS CVn   &   6.4 & G8III &      &  22.42 & 22.4  & 19.7 &  21  & 45 & 0.6 &     & 0.8        & \\
HK Lac       & 209813 &  RS CVn   &   6.6 & K0III &      &  24.15 & 24.4  & 20   &  20  & 65 & 0.7 &     & 0.7        & \\
OU And       & 223460 &  giant    &   7.4 & G0III &      &  24.2  &       & 20   &      &    & 0.6 &     & $\ge$ 0.7  & \\
IM Peg       & 216489 &  RS CVn   &  10.3 & K2III & 4.5 &  24.45 & 24.65 & 28.2 &  28  & 53 & 1.5 & 0.7 & 1.6        & \\
%HR Peg      & 216672 &           &   3.4 & S     & 0.9 &        &       &      &      &    &     &     &           &  \\
KU Peg       & 218153 &  RS CVn   &   5.3 & G8III &      &  25.9  & 25.3  & 23.1 &  28  & 50 & 0.5 &     & 0.8        & \\
II Peg       & 224085 &  RS CVn   &  23.6 & K2V   & 4.6 &   6.73 &  6.72 & 23.1 &  23  & 61 & 0.2 &     & 0.8        & \\ 
%V1321 Ori    &    /   & WTTS      &        & K0III &      &        &  5.68 &      &  71  & 62 &     &     &           &  \\ 
V830 Tau     &    /   & WTTS      &        & M0    & 8.5 &        &  2.75 &      &  34  & 80 &     & 0.1 &            & \\
He 520       &        & alpha Per &  5.5  & G5V   &      &        &  0.61 &      &  91  & 75 &     &     & 0.1        & \\
He 699       &        & alpha Per &  5.5  & G2V   &      &        &  0.49 &      &  97  & 65 &     &     & 0.1        & \\
HII 3163     &        & Pleiad    &  8.5  & K0V   &      &        &  0.42 &      &  70  & 58 &     &     & 0.1        & \\
HII 686      &        & Pleiad    &  8.5  & K4V   &      &        &  0.40 &      &  64  & 51 &     &     & 0.1        & \\
%RXJ1508.6-4423&    & WTTS      &        & G2V   &      &        &  0.31 &      & 115  & 28 &     &     &           &  \\
\hline
\end{tabular}
\end{table*}

\begin{table}[b]
\caption{Estimated diameters for giants with predicted 
surface activity (Choi et al. 1995). 
A  measured diameter is known only for $\Theta^1$\,Tau,
namely 2.6\,mas (White \& Feierman (1987).}
\begin{tabular}{{lrrrlrrrrrrrrrr}}
\hline
Name  & HD   &$\pi$& SpT &  K & P &  $\theta_1$&$\theta_2$ \\
      &            & mas &   & mag & d &  mas  & mas   \\
$\Theta^1$ Tau & 28307 & 20.7 & G9III & 1.6 & 140 & 2.2 & 2.2 \\
$\nu^3$ CMa & 47442  & 7.0 & K1III & 1.9 & 183 & 0.8 & 1.9 \\
19 Pup & 68290  & 17.6 & K0 III & 2.6 & 159 & 1.9 & 1.4 \\
       & 73598  & 4.5  & G8III &  & 159 & 0.4 &  \\
39 CnC & 73665  & 5.6 & G9III & 4.2 & & 0.6 & 0.7 \\
       & 73710  & 4.3 & K0III & 4.2 & 155 & 0.5 & 0.7 \\
       & 73974  & 3.6 & K0III &     & 112 & 0.4 &     \\
$\zeta$ Crt & 102070  & 9.3 & G8III & 2.5 & 137 & 0.9 & 1.5 \\
$\gamma$ Com & 108381  & 19.2 & K2III & 1.9 &     & 2.7 & 2.2 \\
  & 157527  & 10.8 & G7III &  & 147 & 1.0 &  \\
$\iota$ Cap & 203387 & 15.1 & G8III & 2.3 & 68 & 1.4 & 1.7  \\
\hline
\end{tabular}
\end{table}
\end{appendix}
\end{document}